\documentclass[%
 reprint,
superscriptaddress,
 amsmath,amssymb,
 aps,
prx,
]{revtex4-2}
\usepackage{diagbox}
\usepackage{graphicx}% Include figure files
\usepackage{dcolumn}% Align table columns on decimal point
\usepackage{bm}% bold math
\usepackage{hyperref}% add hypertext capabilities
\usepackage{float}
\usepackage{xcolor}

\begin{document}

\preprint{APS/123-QED}

\title{Inhomogeneous Spin excitations in weakly coupled spin-$\frac{1}{2}$ chains}

\author{L. Shen}

 \affiliation{Division of Synchrotron Radiation Research, Lund University, SE-22100 Lund, Sweden}
 
\author{E. Campillo}
 \affiliation{Division of Synchrotron Radiation Research, Lund University, SE-22100 Lund, Sweden}
\author{O. Zaharko}
 \affiliation{Laboratory for Neutron Scattering and Imaging, Paul Scherrer Institut, CH-5232 Villigen PSI, Switzerland}
\author{P. Steffens}
 \affiliation{Institut Laue-Langevin, 71 Avenue des Martyrs, 38042 Grenoble Cedex 9, France}
 \author{M. Boehm}
 \affiliation{Institut Laue-Langevin, 71 Avenue des Martyrs, 38042 Grenoble Cedex 9, France}
 \author{K. Beauvois}
\affiliation{Institut Laue-Langevin, 71 Avenue des Martyrs, 38042 Grenoble Cedex 9, France}
\author{B. Ouladdiaf}
 \affiliation{Institut Laue-Langevin, 71 Avenue des Martyrs, 38042 Grenoble Cedex 9, France} 
 \author{Z. He}
 \affiliation{State Key Laboratory of Structural Chemistry, Fujian Institute of Research on the Structure of Matter, Chinese Academy of Sciences, Fuzhou, Fujian 350002, China}
 \author{D. Prabhakaran}
\affiliation{Department of Physics, University of Oxford, Clarendon Laboratory, Oxford, OX1 3PU, United Kingdom}
\author{A. T. Boothroyd}
 \affiliation{Department of Physics, University of Oxford, Clarendon Laboratory, Oxford, OX1 3PU, United Kingdom} 
\author{E. Blackburn}

 \affiliation{Division of Synchrotron Radiation Research, Lund University, SE-22100 Lund, Sweden}


\begin{abstract}

We present a systematic inelastic neutron scattering and neutron diffraction study on the magnetic structure of the quasi-one-dimensional spin-$\frac{1}{2}$ magnet SrCo$_{2}$V$_{2}$O$_{8}$, where the interchain coupling in the N\'eel-type antiferromagnetic ground state breaks the static spin lattice into two independent domains. At zero magnetic field, we have observed two new spin excitations with small spectral weights inside the gapped region defined by the spinon bound states. In an external magnetic field along the chain axis, the N\'eel order gets partially destabilized at $\mu_\mathrm{0}H^{\star}$~=~2.0~T and completely suppressed at $\mu_\mathrm{0}H_\mathrm{p}$~=~3.9~T, above which a quantum disordered Tomonaga-Luttinger liquid (TLL) prevails. The low-energy spin excitations between $\mu_\mathrm{0}H^{\star}$ and $\mu_\mathrm{0}H_\mathrm{p}$ are not homogeneous, containing the dispersionless (or weakly dispersive) spinon bound states excited in the N\'eel phase and the highly dispersive psinon-antipsinon mode characteristic of a TLL. We propose that the two new modes at zero-field are spinon excitations inside the domain walls. Since they have a smaller gap than those excited in the N\'eel domains, the underlying spin chains enter the TLL state via a local quantum phase transition at $\mu_\mathrm{0}H^{\star}$, making the N\'eel~/~TLL coexistence a stable configuration until the excitation gap in the N\'eel domains closes at $\mu_\mathrm{0}H_\mathrm{p}$.

\end{abstract}

\maketitle

%\tableofcontents

\section{\label{sec:1}Introduction}
At the absolute zero of temperature, a continuous quantum phase transition (QPT) can occur upon variation of a non-thermal control parameter — pressure, chemical substitution, magnetic field and so on — because of the fluctuations inherent in Heisenberg’s uncertainty principle, exposing a singularity called quantum critical point (QCP) that separates the two phases involved \cite{Sachdev2011}. A $\mathit{T}$~$>$~0~K non-thermal phase transition can also be approximated as a QPT as long as the quantum fluctuations overwhelm the thermal fluctuations. This renders experimental assessment of quantum criticality possible \cite{Merchant2014,Coldea_2010}. There are, however, exceptional cases where a QCP is avoided, suppressing the otherwise prominent quantum critical dynamics near the QCP. Notably, masked quantum criticality arises in some of the most studied correlated electron systems, including the heavy-fermion metals \cite{Mathur_1998,Yuan_2003}, itinerant-electron magnets \cite{Pfleiderer_2004,Uemura_2007,Friedemann_2018}, and arguably also the high-temperature superconductors \cite{Broun_2008,Shibauchi_2014}. As a result, classification of the phase formation in the vicinity of a QCP is at the heart of understanding the rich complexity in these materials.

A novel magnetic-field-induced QPT has been predicted to occur in the one-dimensional (1D) spin-$\frac{1}{2}$ Heisenberg-Ising XXZ model \cite{Yang_1966,Haldane_1980,Bogoliubov_1986},
\begin{equation} \label{eq:XXZ}
\begin{aligned}
    H_{XXZ} =&\ J \sum_{i} (S_{i}^{x}S_{i+1}^{x}+S_{i}^{y}S_{i+1}^{y} + \Delta S_{i}^{z}S_{i+1}^{z}) \\
    &\,\,\,\, - g_z\mu_{0}\mu_{B}H\sum_{i}S_{i}^{z},
\end{aligned}
\end{equation}
where $\it{J}$ is the nearest-neighbor antiferromagnetic (AFM) intrachain exchange constant, $\Delta$ is the anisotropy parameter and $g_{z}$ is the component of the Land\'e g-tensor along the chain direction ($\it{z}$~axis). In the Heisenberg-Ising regime ($\Delta$~$>$~1), this model has a gapped AFM ground state of N\'eel type with the spins lying along the z axis [Fig.~\ref{fig:1}(a)]. An external magnetic field ($\mathbf{H}$~//~$\it{z}$~axis) tunes the density of the elementatry excitations (spinons \cite{Faddeev_1981}), acting like a chemical potential to close the gap in the excitation spectrum at a QCP that separates the N\'eel state from a quantum disordered Tomonaga-Luttinger liquid (TLL). A similar concept has been extensively explored in the dimerized antiferromagnets, where the QPT is the Bose-Einstein condensation (BEC) of triplons \cite{Giamarchi_2008,Zapf_2014}.

Recently, the quasi 1D quantum magnet family $\it{M}$Co$_{2}$V$_{2}$O$_{8}$ ($\it{M}$~=~Sr,~Ba), where the intrachain spin-spin interactions can be approximated by the Heisenberg-Ising XXZ model (Eq.~\ref{eq:XXZ}), has attracted tremendous attention \cite{Kimura_2008,Canevet_2013,Grenier_2015,Wang_2016,Bera_2017,Wang_2018,Wang_2019,Faure_2019,Shen_2019,Bera_2020}. While a magnetic-field-induced N\'eel to TLL phase transition has been observed at low temperatures, the magnetic structure in the QPT region has never been evaluated. This issue is important for understanding the quantum criticality in this material because Eq.~\ref{eq:XXZ} is not a comprehensive description for the interactions in materials of this class. In particular, it does not account for the effect of dimensional crossover that becomes energetically relevant while approaching $\mathit{T}$~=~0~K. In this study, we used neutron diffraction and inelastic neutron scattering to measure the magnetic structure of SrCo$_{2}$V$_{2}$O$_{8}$ in the zero-field N\'eel state and across the field-induced N\'eel to TLL QPT. The spinon excitations in both regions are found to be inhomogeneous. Earlier neutron diffraction work has shown that in the N\'eel phase, there are two magnetic domains with approximately equal populations \cite{Shen_2019}. We present a model that takes this into account to interpret our observations.

\section{\label{sec:2}Methods}

Two inelastic neutron scattering (INS) experiments were performed on the cold-neutron triple-axis spectrometer, ThALES, at the Institut Laue-Langevin (ILL). A cylindrical SrCo$_{2}$V$_{2}$O$_{8}$ single crystal (height: 22 mm, diameter 6 mm, mass: 2.699 g) grown in Oxford (UK) by the floating zone method \cite{Lejay_2011} was used. In the first experiment \cite{ILLDATA1}, it was mounted in a dilution refrigerator inside a 6 T horizontal cryomagnet. The initial and final neutron wavevectors were selected using a Si (111) monochromator and analyzer, respectively. The INS spectra at momentum transfers with a finite out-of-plane component were collected in this setup, with a fixed final wavevector of 1.5~$\mathrm{\AA}^{-1}$. In the second experiment \cite{ILLDATA2}, it was mounted in a dilution refrigerator inside a 10 T vertical cryomagnet. The initial and final neutron wavevectors were selected using a PG (002) monochromator and analyzer, respectively. The INS spectra in the reciprocal (H, K, 0) plane were collected in this setup, with a fixed final wavevector of 1.3~$\mathrm{\AA}^{-1}$. All of the INS data presented here were measured with the $\bm{c}$ axis aligned along the magnetic field.

The single crystal neutron diffraction measurements for the magnetic structure refinement at 3.0~T and 50~mK were performed on the thermal neutron diffractometer ZEBRA at the Swiss Spallation Neutron Source (SINQ) at the Paul Scherrer Institute. A SrCo$_{2}$V$_{2}$O$_{8}$ single crystal ($\sim$~3~$\times$~3~$\times$~6~mm$^{3}$) grown in Fujian (China) by the spontaneous nucleation method \cite{He_2006} was measured with a lifted detector (normal beam geometry). It was mounted in a dilution refrigerator inside a 6~T vertical cryomagnet, with the $\bm{c}$ axis aligned along the magnetic field. For the magnetic structure determination, a set of magnetic and nuclear reflections were collected at neutron wavelength $\lambda$~=~1.178~$\mathrm{\AA}$ using a Ge (311) monochromator. We have also performed additional measurements on the same crystal on the D10 diffractometer (ILL) \cite{ILLDATA3} in order to search for the possible emergent spin modulation between 2.0~T and 3.9~T (see Section \ref{sec:5} and Supplemental Material). The sample was mounted in a dilution refrigerator inside a 6~T horizontal cryomagnet. A vertically focusing graphite monochromator was used, fixing the wavelength of the incoming neutrons to $\lambda$~=~2.36~$\mathrm{\AA}$.

The quality of the two single crystals was checked by neutron Laue diffraction; no impurity could be resolved. In addition, they share identical magnetic field versus temperature phase diagram within the detection limit [see Fig.~\ref{fig:S1}(a) in the Supplemental Material]. Unless otherwise stated, the data presented in the main text below are based on the measurements on the sample grown by floating zone \cite{Lejay_2011}. 

\section{\label{sec:3}Magnetic dimensional crossover in SrCo$_{2}$V$_{2}$O$_{8}$}

In this section, we discuss the impact of dimensional crossover on the spin excitations based on a minimal model that approximates the interchain exchange interactions in SrCo$_{2}$V$_{2}$O$_{8}$. This model will be used to understand the experimental observations presented in the following sections.

\begin{figure}[b]
\centering
	\includegraphics[width=0.49\textwidth]{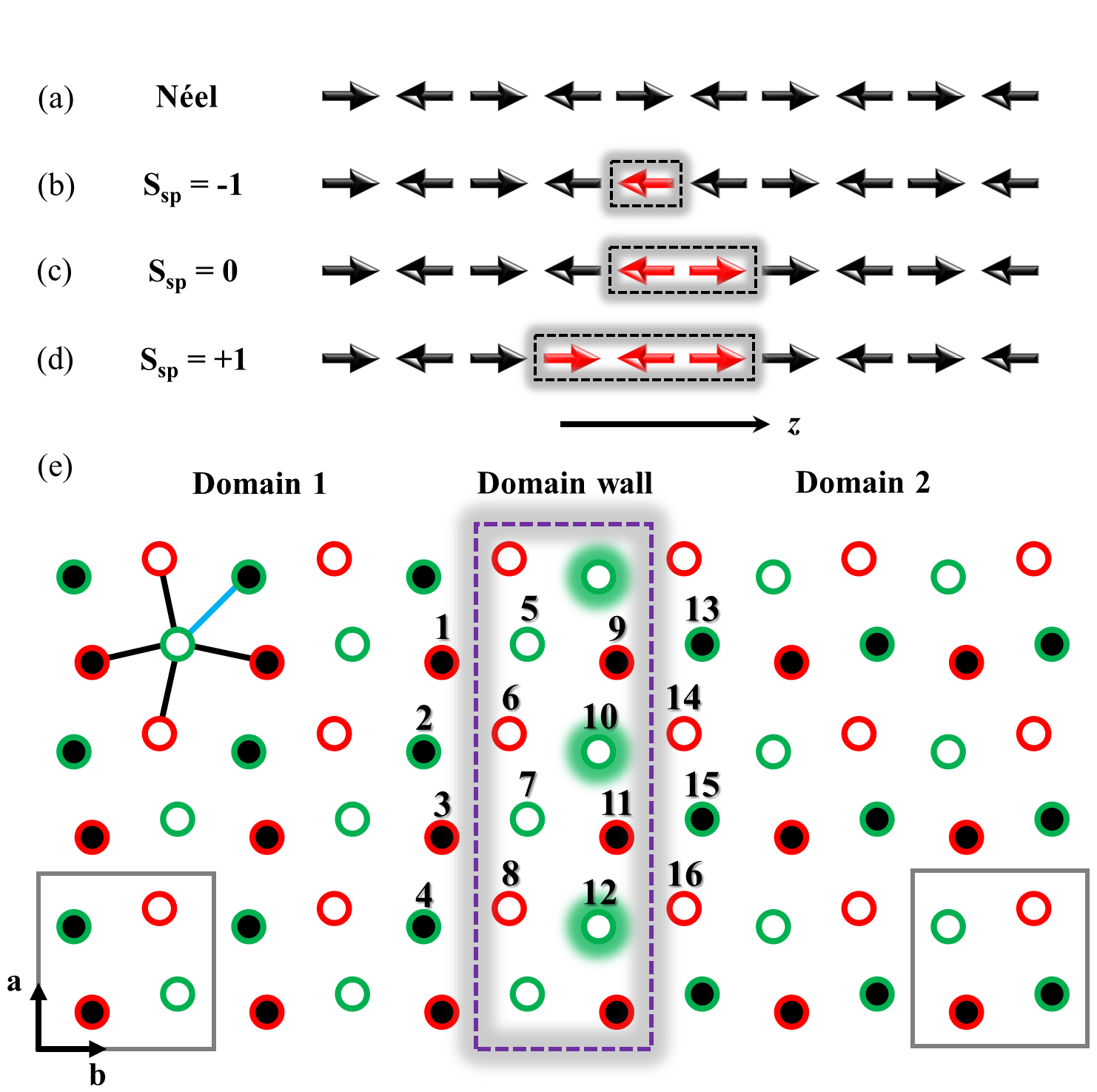}
    \caption{Magnetic structure and spinon excitations. (a) N\'eel-type AFM order predicted by Eq.~\ref{eq:XXZ}. (b-d) Spinon pairs created by spin flips (red arrows), with $S_\mathrm{sp}$~=~-1, 0, +1, respectively. (e) In-plane magnetic structure of SrCo$_{2}$V$_{2}$O$_{8}$ at zero-field. There are two independent Co spin screw chains denoted by the green and red circles \cite{Shen_2019}. The filling colors (black or white) denote spin direction (up or down). Gray boxes enclose inequivalent magnetic unit cells in the two domains present, which are separated by the domain wall enclosed by the dashed purple box.  The solid lines between Co sites denote the different interchain exchange paths that contribute to the local molecular field $h_s$ at each site (blue = $\pm{}h_1$, black = $\pm{}h_2$) . The glowing circles highlight spins that have been flipped by domain wall formation. Sixteen spin sites are labelled for the $|h_s|$ analysis in Table~\ref{Tab:1}.
    }
\label{fig:1}
\end{figure}

The elementary excitations in a spin-$\frac{1}{2}$ chain are spin-$\frac{1}{2}$ spinons \cite{Faddeev_1981}, which can be thought of as quasiparticles that are always created or annihilated in pairs by spin flips. Consequently, each spinon pair has a bosonic quantum spin number of $S_\mathrm{sp}$~=~+1, -1 or 0, depending on the number and direction of spin flips [Fig.~\ref{fig:1}(b-d)]. Fundamentally, the zero-field excitation gap in the N\'eel state is determined by the doubly degenerate $S_\mathrm{sp}$~=~$\pm$1 spinon pairs; the condensation of the $S_\mathrm{sp}$~=~+1 branch at the QCP drives the N\'eel to TLL QPT in Eq.~\ref{eq:XXZ}. As a result, the spinon dynamics are pivotal to the location of a QCP.

The dimensional crossover brought about by lowering the temperature in a quasi 1D magnet introduces additional interchain couplings. Unlike the intrachain counterparts, these terms are perturbative and can be treated in a mean-field approximation \cite{Shiba_1980,Nagler_1983,Mena_2020}. Specifically, their contributions can be modelled by adding an effective staggered magnetic field $h_s$(-1)$^{\mathit{i}}$ to the $i^{th}$ spin in a single XXZ chain (Eq.~\ref{eq:XXZ}). In the absence of interchain couplings, it costs the same energy to flip arbitrary number of spins, meaning that spinon pairs propagate freely along the chain, giving a continuous excitation spectrum. When $h_s$ is non-zero, flipping $\it{n}$ spins in one chain leads to an additional energy gain that is proportional to $\it{n}${}$h_s$. This is equivalent to putting the spinon pairs in an attractive potential, confining them spatially \cite{Lake_2010}. Confinement replaces the free spinon continuum with a series of discrete modes, i.e.~spinon bound states \cite{Shiba_1980}. In the $\it{M}$Co$_{2}$V$_{2}$O$_{8}$ ($\it{M}$~=~Sr,~Ba) family, this confinement is well described by a 1D Schr\"odinger equation with linear potential \cite{Grenier_2015,Wang_2016,Bera_2017}, the solutions of which are:
\begin{equation}\label{LinConf}
    E_{j}=2E_{\mathrm{0}}+\zeta_{j}h_\mathrm{s}^{\frac{2}{3}}(\frac{\hbar^{2}}{\mu})^{\frac{1}{3}}, (j = 1, 2, 3...),
\end{equation}
where $E_{\mathrm{0}}$ is the free spinon excitation gap, $E_{j}$ is the energy of the $j^{th}$ bound state, $\zeta_{j}$ is the $j^{th}$ negative zero of the Airy function, and $\mu$ is the reduced spinon mass \cite{McCoy_1978a,McCoy_1978b}. Based on this equation, the zero-field $S_\mathrm{sp}$~=~$\pm$1 spinon excitation gap $E_{1}$ is proportional to $h_\mathrm{s}$.

SrCo$_{2}$V$_{2}$O$_{8}$ possesses screw chains of Co$^{2+}$ ions with effective spin $\frac{1}{2}$ running along the crystallographic $\bm{c}$ axis. The validity of the XXZ model (Eq.~\ref{eq:XXZ}) in describing the intrachain interactions in this system has been confirmed by multiple experimental studies \cite{Bera_2017,Wang_2018,Bera_2020}. Upon cooling, the Co$^{2+}$ spins develop a three-dimensionally ordered AFM structure at $T_\mathrm{N}$~=~5.0~K \cite{Shen_2019}. In the ground state, the spin configuration within each chain agrees well with the easy-axis N\'eel order depicted in Fig.~\ref{fig:1}(a). The relative spin alignment between two neighbouring chains is not trivial. This is due to the symmetry reduction associated with the bulk AFM phase transition at $T_\mathrm{N}$, which decouples half of the spin chains from the rest. The resulting spin arrangements in a given $ab$ plane are shown in Fig.~\ref{fig:1}(e). Each in-plane magnetic unit cell hosts two diagonal AFM spin pairs that are independent of each other. This generates two types of domains, distinguished by their magnetic unit cell configurations, which are almost equally populated in space \cite{Shen_2019}. We note that this two-domain configuration in SrCo$_{2}$V$_{2}$O$_{8}$ is exclusively caused by the symmetry reduction in the N\'eel state, not by the minimization of magnetostatic energy. The latter is responsible for the formation of conventional domains.

The inhomogeneous spin lattice discussed above is not thermally activated and therefore can persist down to zero temperature. We now analyze its influence on the spinon excitations. Based on a structural analysis \cite{Bera_2015}, the interchain exchange network in SrCo$_{2}$V$_{2}$O$_{8}$ contains four independent paths: two in the skewing direction and the other two in the $ab$ plane. For simplicity, we only consider the two in-plane interchain couplings, the effective molecular fields of which are denoted as $\pm{}h_1$ and $\pm{}h_2$. Their signs depend on whether the relevant spin pairs are antiferromagnetically (minus) or ferromagnetically (plus) aligned. For a given Co site, five neighboring spins are involved [Fig.~\ref{fig:1}(e)]. The total effective staggered magnetic field $h_s$ is obtained by summing up these individual molecular fields \cite{Nagler_1983,Mena_2020}.

We have calculated the $|h_s|$ values for the 16 spin sites labelled in Fig.~\ref{fig:1}(e); their spatial distribution is summarized in Table~\ref{Tab:1}. For any given spin inside the domains, e.g. spin $\#$~1-4, $|h_s|$ is always equal to $h_1$. However, inside the domain walls, which are formed by planes of spin chains in three dimensions, $h_{s}$ is nonuniform.  For example, for spin $\#$~5 it is $h_1$, for spin $\#$~6 it is $|h_1\,-\,2h_2|$, and for spin $\#$~9 it is $|h_1\,+\,2h_2|$. Moreover, even after the two skewing paths are included for a comprehensive analysis on the interchain exchange network \cite{Bera_2015}, we find that the nonuniformity of $h_{s}$ is still discrete. In other words, there exists well defined spinon bound states corresponding to a finite set of $|h_s|$ values \cite{Nagler_1983}. Accordingly, our simplified model captures the essential information: the spinon excitations in the domain walls have distinct gap values (Eq.~\ref{LinConf}) from those in the domains because of variations in the confining potential.

\begin{table}[H]
  \centering
  \caption{Spatial distribution of the effective staggered magnetic field $h_s$ in the bulk N\'eel state. Only the in-plane interchain couplings marked in Fig.~\ref{fig:1}(e) are considered.}
  \label{Tab:1}
  \begin{ruledtabular}
  \begin{tabular}{c|ccc}
$|h_s|$&$|h_1|$&$|h_1\,-\,2h_2|$&$|h_1\,+\,2h_2|$\\
  \hline
 Site label&1-4, 5, 7, 10, 12, 13-16& 6, 8 & 9, 11
  \end{tabular}
  \end{ruledtabular}
\end{table}

\section{\label{sec:4}Results}

\begin{figure*}
\centering
	\includegraphics[width=0.8\textwidth]{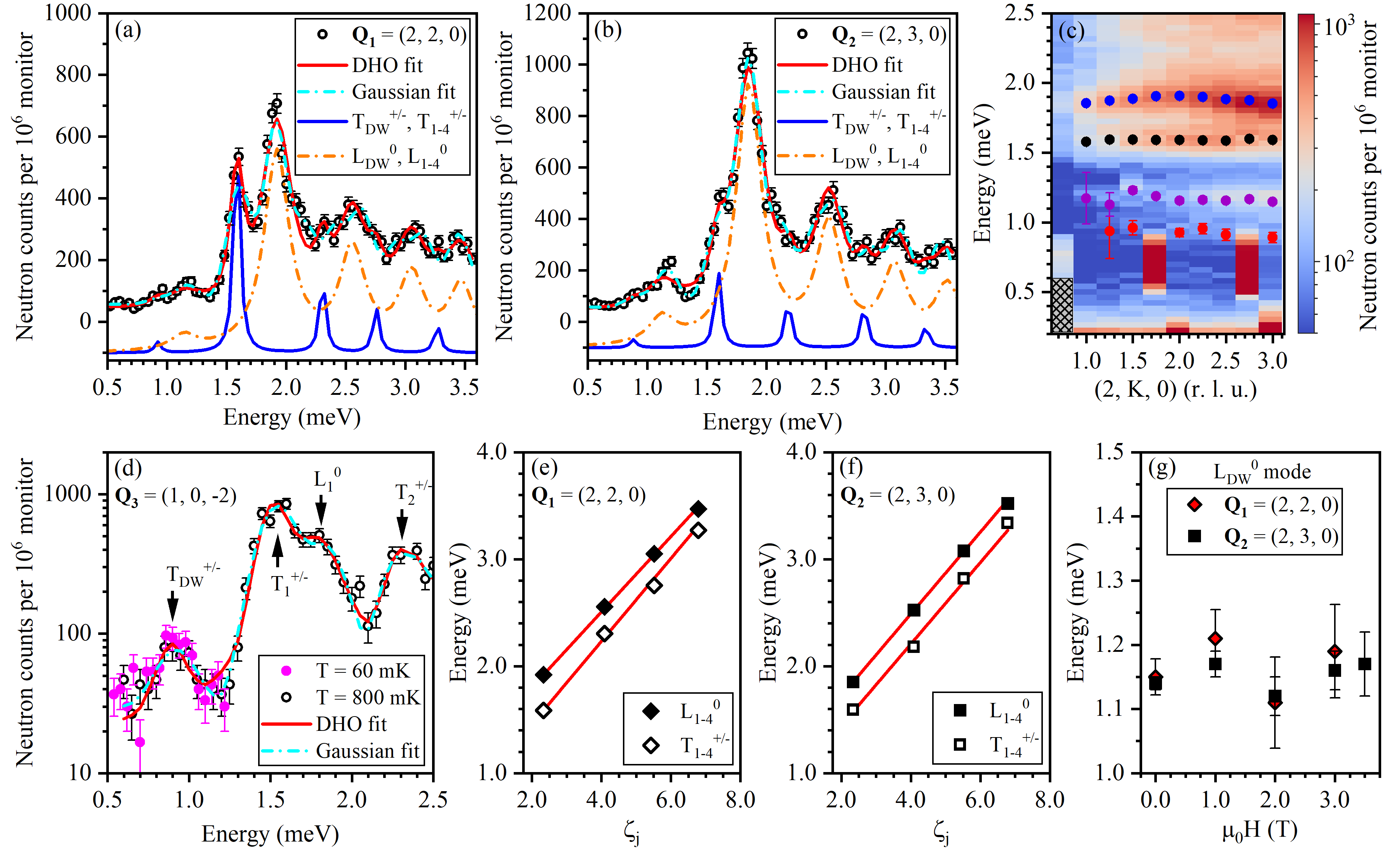}
    \caption{Spinon confinement in SrCo$_{2}$V$_{2}$O$_{8}$. (a-c) Spinon bound states at 50~mK measured by inelastic neutron scattering at the momentum transfers $\mathbf{Q_{1}}$~=~(2,~2,~0) and $\mathbf{Q_{2}}$~=~(2,~3,~0), and their dispersion in the reciprocal (2,~K,~0) plane. In the $\mathrm{T_\mathrm{DW}^{+/-}}$~/~$\mathrm{T_{1-4}^{+/-}}$ (blue solid line) and $\mathrm{L_\mathrm{DW}^{0}}$~/~$\mathrm{L_{1-4}^{0}}$ (orange dash-dotted line) plots, the instrumental resolution profile has been deconvolved to show the intrinsic damping of these modes. The $\mathrm{T_\mathrm{DW}^{+/-}}$ (red circles), $\mathrm{L_\mathrm{DW}^{0}}$ (purple circles), $\mathrm{T_{1}^{+/-}}$ (black circles) and $\mathrm{L_{1}^{0}}$ (blue circles) mode energies are plotted in (c), where the `bright' areas around 0.75~meV and at K~=~0.75,~1.75 and 2.75~r.~l.~u. are spurious leaks through the analyzer; they were excluded in our data analysis. (d) Spinon bound states at 60~mK (800~mK) and $\mathbf{Q_{3}}$~=~(1,~0,~2), displayed on a logarithmic scale to highlight the $\mathrm{T_\mathrm{DW}^{+/-}}$ mode. (e, f) Comparing the central energies of the domain spinon bound states with the linear confinement theory described by Eq.~\ref{LinConf}. (g) Magnetic field dependence of the $\mathrm{L_\mathrm{DW}^{0}}$ mode energy at $\mathbf{Q_{1}}$ and $\mathbf{Q_{2}}$. The numerical fits are described in the main text. }
\label{fig:2}
\end{figure*}

\subsection{Domain wall spinon excitations in SrCo$_{2}$V$_{2}$O$_{8}$}

In Fig.~\ref{fig:2}, INS spectra as a function of momentum transfer $\mathbf{Q}$ and neutron energy loss $E$ are presented.  As expected \cite{Shiba_1980}, a series of discrete modes - the spinon bound states - are seen.  Figure \ref{fig:2}(a)-(c) shows the behaviour at the in-plane nuclear zone center $\mathbf{Q_{1}}$~=~(2,~2,~0), the AFM zone center (H~+~K~+~L~=~odd integer) $\mathbf{Q_{2}}$~=~(2,~3,~0), and a dispersion map of the magnetic excitations in the reciprocal (2,~K,~0) plane. A series of peaks are observed. To quantitatively analyze our data, we fit them numerically. The instrumental resolution function, which has a full width at half maximum (FWHM) of 88(3)~$\mu$eV, has been deconvolved. We have also evaluated the detection noise by conducting measurements at several momentum transfer points where there are negligible sample contributions; these calibrations show that the background may be modelled as a constant term. In the literature, a Gaussian model, where all peaks share a common width, has been used to describe the lineshape of the spinon bound states \cite{Grenier_2015,Bera_2017,Bera_2020}. For our high energy resolution data, we find that modelling the peaks with a damped harmonic oscillator (DHO), the width of which depends on whether the corresponding spinon bound state has $S_\mathrm{sp}$~=~$\pm$1 or 0, provides considerably better fitting quality [Fig.~\ref{fig:2}(a)~$\&$~(b)]. This high energy resolution is necessary to resolve these subtle features because once we relax the resolution FWHM to 160(8)~$\mu$eV, the DHO / 2-width and Gaussian / 1-width models produce similar fitting quality; a typical example is demonstrated in Fig.~\ref{fig:2}(d) for the spectrum collected at another AFM zone center $\mathbf{Q_{3}}$~=~(1,~0,~2). 

In Fig.~\ref{fig:2}(a)~$\&$~(b), the four pairs of spinon bound states above 1.40~meV with strong spectral weights, which are excited inside the N\'eel domains, have been reported in the literature \cite{Bera_2020,Wang_2018,Bera_2017}. The $S_\mathrm{sp}$~=~0 and $\pm$1 modes account for the magnetic moment fluctuations along the $\bm{c}$ axis (longitudinal, or $L$, mode) and in the $ab$ plane (transverse, or $T$, modes), respectively \cite{Affleck_1992,Schulz_1996}. Since neutron scattering is only sensitive to the spin configuration perpendicular to $\mathbf{Q}$ and the N\'eel order is dominated by the Ising-like easy axis anisotropy along the $\bm{c}$ axis \cite{Shen_2019}, the four strong peaks (orange dash-dotted line) detected in this in-plane geometry are the $S_\mathrm{sp}$~=~0 spinon bound states, while the four weaker ones (blue solid line) are the $S_\mathrm{sp}$~=~$\pm$1 spinon bound states. These modes are either weakly dispersive with a bandwidth of about 0.05~meV or dispersionless, supporting the hypothesis of insignificant interchain spinon correlations. 

We also observe a previously unreported pair of peaks with small spectral weights at about 0.88~meV and 1.15~meV [Fig.~\ref{fig:2}(a)-(c)]. Their central energies lie in the gapped region of the aforementioned bulk spinon spectrum. Their evolution in the $\mathbf{Q}$-E space is reminiscent of the other spinon modes at higher energies: (1) they are non-dispersive in the transverse direction, and (2) the peak at 0.88~meV is less intense than that at 1.15~meV. We interpret these peaks as another $S_\mathrm{sp}$~=~$\pm$1 ($\sim$~0.88~meV) and 0 ($\sim$~1.15~meV) pair. In Fig.~\ref{fig:2}(g) we track the $\sim$ 1.15~meV mode as a function of magnetic field, and see no shift, indicating that it behaves like a $S_\mathrm{sp}$~=~0 longitudinal mode.  Due to the small spectral weight we were not able to track the $\sim$ 0.88~meV mode in a magnetic field; if it is an $S_\mathrm{sp}$~=~$\pm$1 mode, it should show a Zeeman splitting. As an alternate way of testing this assignment, we also measured the same energy range at $\mathbf{Q_{3}}$~=~(1,~0,~2).  In this out-of-plane geometry, we should be more sensitive to the $S_\mathrm{sp}$~=~$\pm$1 modes than to the $S_\mathrm{sp}$~=~0 mode \cite{Grenier_2015,Bera_2017}. As Fig.~\ref{fig:2}(d) shows, this is exactly what is seen, both for the new peaks at 0.88~meV and 1.15~meV (below the detection limit in this geometry), and for the well-studied modes between 1.40~meV and 1.95~meV. This measurement strongly supports the $S_\mathrm{sp}$~=~$\pm$1 nature of the peak at 0.88~meV.

Trivial explanations for the two weak peaks below 1.40~meV include crystalline mosaicity and impurity. The first can be easily ruled out because the zone centers probed ($\mathbf{Q_{1}}$, $\mathbf{Q_{2}}$ and $\mathbf{Q_{3}}$) correspond to the energy minima of the spinon spectrum \cite{Bera_2017,Bera_2020}. In the absence of strong in-plane dispersion [Fig.~\ref{fig:2}(c)] \cite{Bera_2017}, any secondary misaligned crystal of SrCo$_{2}$V$_{2}$O$_{8}$ can only add spectral weights to higher energies. On the other hand, given that only a few compounds have been experimentally confirmed to host a well-defined $S_\mathrm{sp}$~=~0 mode \cite{Grenier_2015,Bera_2017}, it is very difficult to reconcile with the presence of such mode in a random impurity phase.

In theory, the lowest-lying spinon bound state should mostly consist of one-spin flips [Fig.~\ref{fig:1}(b)], and should always have the largest spectral weight \cite{Shiba_1980,Bera_2017}. Accordingly, the peaks at 0.88~meV and 1.15~meV do not come from the bulk N\'eel domains, which give rise to the stronger peaks at higher energies \cite{Bera_2017}. Following the discussion in Section~\ref{sec:3}, and given that INS is a powerful technique for detecting soliton-like excitations with a small volume fraction \cite{Kenzelmannn_2003,Haravifard_2006}, we therefore propose that they are in fact the lowest-lying spinon bound states in the domain walls. This naturally explains their much smaller spectral weights comparing to those excited in the domains, as the domain walls occupy only a small portion of the sample. One necessary condition for any spinon bound state to rise in the sample is that it must be above the confinement threshold 2$E_{0}$ (Eq.~\ref{LinConf}). We have calculated this parameter for different $S_\mathrm{sp}$s in SrCo$_{2}$V$_{2}$O$_{8}$ using the spinon bound state energies in the domains. As shown in Fig.~\ref{fig:2}(e)~$\&$~(f), the linear confinement theory fits the data well. We obtain 2$E_{\mathrm{0}}$~$\simeq$~0.68~meV for $S_\mathrm{sp}$~=~$\pm$1 and 2$E_0$~$\simeq$~1.02~meV for $S_\mathrm{sp}$~=~0. Based on this calculation, the necessary condition for spinon confinement in the domain walls is fulfilled. According to Eq.~\ref{LinConf} and Table~\ref{Tab:1}, there exists multiple sets of domain wall spinon bound states above 2$E_{\mathrm{0}}$ due to the nonuniformity of $h_{s}$. However, experimental detection of any of these modes above 1.4~meV is challenging due to the small spectral weights and dominant contributions from domain spinons.

We will now use the labels $\mathrm{T}_{i}^{+/-}$ or $\mathrm{L}_{i}^{0}$ ($\it{i}$~=~1-4) to denote the $i^{th}$ $S_\mathrm{sp}$~=~$\pm$1 or 0 spinon bound state in the domains, and $\mathrm{T}_\mathrm{DW}^{+/-}$ or $\mathrm{L}_\mathrm{DW}^{0}$ to denote the observed $S_\mathrm{sp}$~=~$\pm$1 or 0 spinon bound state in the domain walls. The high-energy-resolution INS spectroscopy data allow us to conclude that the zero-field spinon confinement in SrCo$_{2}$V$_{2}$O$_{8}$ is not perfect [Fig.~\ref{fig:2}(a)~$\&$~(b)]. The inverse lifetimes of the $S_\mathrm{sp}$~=~$\pm$1 and $S_\mathrm{sp}$~=~0 modes extracted from the numerical fits are $\mathbf{Q}$-independent within the errors, with FWHMs of 0.06(2)~meV and 0.25(2)~meV, respectively. 

\subsection{Coexistence of N\'eel antiferromagnetism and a Tomonaga-Luttinger liquid}

\begin{figure}[b]
\centering
	\includegraphics[width=0.49\textwidth]{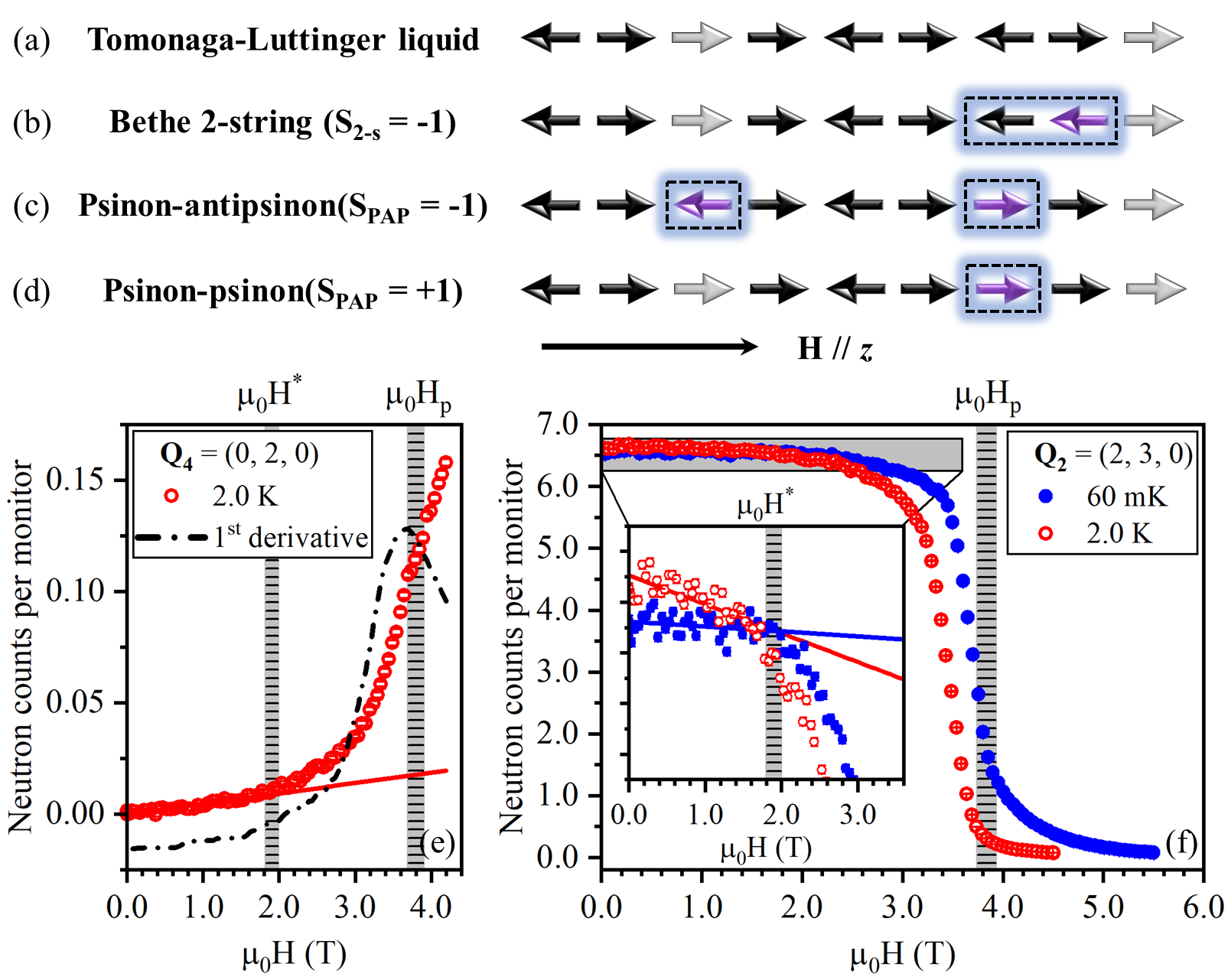}
    \caption{Spin configurations and static magnetic properties in a field-induced Tomonaga-Luttinger liquid (TLL). (a-d) Schematic illustrations of the ground state, Bethe 2-string (2-s), psinon-antipsinon (PAP) and psinon-psinon (PP) excitations of a TLL \cite{Wang_2018,Bera_2020,Yang_2019}. Gray arrows represent the condensed $S_\mathrm{sp}$~=~+1 spinon pairs in a longitudinal magnetic field. Purple arrows are spin flips responsible for the excitations. Magnetic field dependences of the neutron diffraction intensity at (e) $\mathbf{Q_{4}}$~=~(0,~2,~0) and (f) $\mathbf{Q_{2}}$~=~(2,~3,~0), which probe the staggered ferromagnetic moment ($M_{c}$) along the chain and N\'eel order, respectively \cite{Shen_2019}. The nuclear structure contribution in (e) has been subtracted. The inset of (f) is an enlarged view of the shaded area in the main panel. The hatched columns mark the positions of $\mu_\mathrm{0}H^{\star}$ and $\mu_\mathrm{0}H_\mathrm{p}$. The dash-dotted line in (e) is the first derivative of the data (red dots). The solid lines in (e-f) are linear fits to the data points below $\mu_\mathrm{0}H^{\star}$. }
\label{fig:3}
\end{figure}

\begin{figure*}
\centering
	\includegraphics[width=0.58\textwidth]{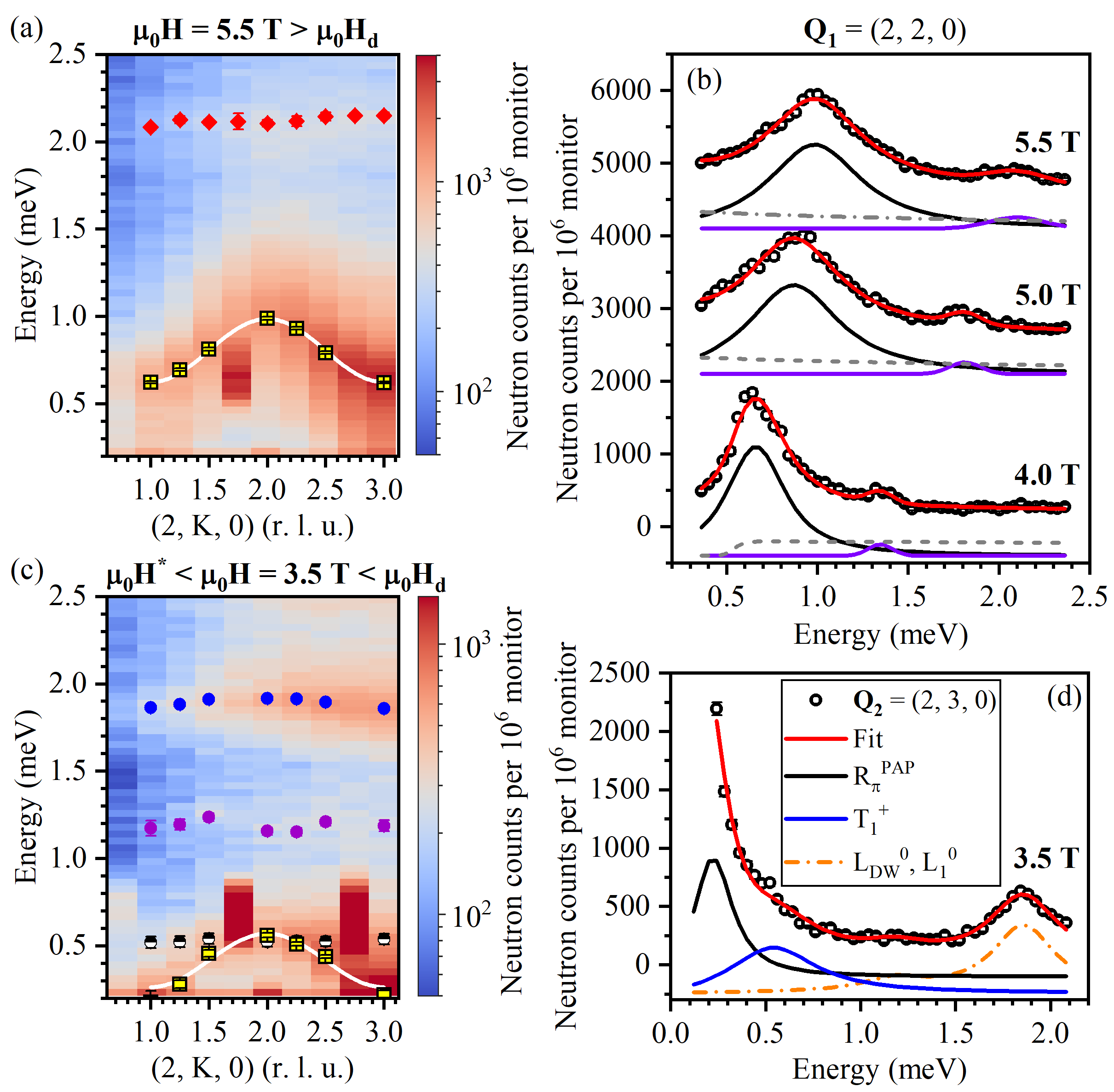}
    \caption{Magnetic excitations in a longitudinal magnetic field. (a) Dispersion of the Tomonaga-Luttinger liquid spin dynamics in the reciprocal (2, K, 0) plane at 5.5 T. The red diamonds and yellow squares mark the positions of the Bethe 2-strings ($\chi_{\pi}^{2-s}$) and psinon-antipsinons ($\mathrm{R_{\pi}^{PAP}}$), respectively. (b) Magnetic excitation spectrum at $\mathbf{Q_{1}}$~=~(2,~2,~0) as a function of magnetic field. The black solid, gray dashed, and violet solid lines are the numerical fits to the $\mathrm{R_{\pi}^{PAP}}$, continuum (see the main text) and $\chi_{\pi}^{2-s}$ modes, respectively. (c) Dispersion of the magnetic excitations at 3.5~T, which shows the coexistence of spinon bound states [$\mathrm{T_{1}^{+}}$ (black circles), $\mathrm{L_\mathrm{DW}^{0}}$ (purple solids) and $\mathrm{L_{1}^{0}}$ (blue solids)] and $\mathrm{R_{\pi}^{PAP}}$ (yellow squares). (d) Magnetic excitation spectrum at $\mathbf{Q_{2}}$~=~(2,~3,~0) and 3.5~T. The white solid lines in (a) and (c) are sinusoidal fits. The numerical fits in (b) and (d) are described in the main text. The spectral weights at low energies in (b) and (d) contain the contributions from elastic scattering; its lineshape has been omitted for clarity. All measurements were carried out at 50~mK.}
\label{fig:4}
\end{figure*}

We study the magnetic structure as a function of longitudinal magnetic field. The onset of the TLL state can be understood as the condensation of the spin flips responsible for the $S_\mathrm{sp}$~=~+1 spinon pairs [see Fig.~\ref{fig:3}(a)]. This is induced by the Zeeman splitting of the transverse spinon mode in a longitudinal magnetic field, and should lead to the development of a staggered ferromagnetic moment, $M_c$, along the $\bm{c}$ axis. We have studied this by measuring the elastic scattering signal, using the energy analysis available on the ThALES three-axis cold-neutron spectrometer. Figure \ref{fig:3}(e) shows the intensity of the Bragg reflection $\mathbf{Q_{4}}$~=~(0,~2,~0) as a function of magnetic field. $\mathbf{Q_{4}}$ is a nuclear zone center, as well as a ferromagnetic zone center. In Fig.~\ref{fig:3}(e) the zero-field nuclear contribution has been subtracted to give a clean probe of $M_{c}$ \cite{Shen_2019,Canevet_2013} as a function of magnetic field.  In the low field region, the intensity scales almost linearly with magnetic field, deviating upwards from $\mu_\mathrm{0}H^{\star}$~=~1.8~T.

Figure~\ref{fig:3}(f) shows similar measurements taken at an antiferromagnetic zone center, in this case $\mathbf{Q_{2}}$~=~(2,~3,~0).  At 2.0~K and below $\mu_\mathrm{0}H^{\star}$, there is a tiny linear decrease in the diffraction intensity at $\mathbf{Q_{2}}$; this effect is almost absent at 60~mK.  These initial changes are most likely driven by thermal fluctuations. We also observe that the N\'eel order does not vanish until $\mu_\mathrm{0}H_{p}$ ($\sim$ 3.75~T at 2.0~K); this is the field above which SrCo$_{2}$V$_{2}$O$_{8}$ becomes disordered at 1.5~K~$<$~$\it{T}$~$<$~$T_{\mathrm{N}}$. Below 1.5~K, $\mu_\mathrm{0}H_\mathrm{p}$ becomes temperature independent and sits at 3.9~T. In this temperature range, a spin density wave develops above $\mu_\mathrm{0}H_\mathrm{p}$, splitting the magnetic reflection at the AFM zone centers, e.g.~$\mathbf{Q_{2}}$, into a pair of peaks along the reciprocal $\bm{c}^{\star}$ axis. The tails of these incommensurate reflections are responsible for the remnant intensities at 60~mK and above $\mu_\mathrm{0}H_\mathrm{p}$ [Fig.~\ref{fig:3}(f) and Fig.~\ref{fig:S1}(b) in the Supplemental Material). These observations are in agreement with our previous study \cite{Shen_2019}.

In the intermediate field region, our measurements detect a response of the $\mathbf{Q_{2}}$ reflection to $\mu_\mathrm{0}H^{\star}$, above which the intensity loss accelerates [inset of Fig.~\ref{fig:3}(f)]. The $\mu_\mathrm{0}H^{\star}$ value also slightly increases to 2.0~T as the temperature is lowered to 60~mK. Since the N\'eel order in SrCo$_{2}$V$_{2}$O$_{8}$ involves two unequally populated magnetic domains \cite{Shen_2019}, the suppression of the $\mathbf{Q_{2}}$ reflection could be caused by a change in the domain volume fractions. A misalignment between the magnetic field and $\bm{c}$ axis \cite{Canevet_2013} or equilibration difficulty at very low temperatures could also be responsible for such changes. We dismiss this possibility based on magnetic structure refinements carried out on the sample grown by spontaneous nucleation \cite{He_2006} at 3.0~T used in Ref.~\cite{Shen_2019}, using 41 magnetic reflections belonging to the N\'eel phase (see Fig.~\ref{fig:S2} in the Supplemental Material). These refinements reveal that, besides a reduction in the ordered magnetic moment at each Co site, both the magnetic structure and volume fractions of the two domains remain unchanged relative to those measured in zero field. This rules out the domain scenario in this sample and confirms that the bulk averaged N\'eel order is partially suppressed between $\mu_\mathrm{0}H^{\star}$ and $\mu_\mathrm{0}H_\mathrm{p}$. 

The kink at $\mu_\mathrm{0}H^{\star}$ revealed in Fig.~\ref{fig:3}(e) supports the emergence of spin chains with a finite $M_{c}$ before the N\'eel order is completely suppressed at $\mu_\mathrm{0}H_\mathrm{p}$. While a finite $M_{c}$ fits the description for having a TLL above $\mu_\mathrm{0}H^{\star}$, it is $\mathit{not}$ direct evidence. Indeed, since the slope of the diffraction intensity at $\mathbf{Q_{4}}$ peaks at $\mu_\mathrm{0}H_\mathrm{p}$ [Fig.~\ref{fig:3}(e)], which probes the suppression of the bulk N\'eel order [Fig.~\ref{fig:3}(f)], $\mu_\mathrm{0}H^{\star}$ could simply mark the starting point of a second-order phase transition, or some sort of crossover behaviour. However, if we assume that a TLL emerges at $\mu_\mathrm{0}H^{\star}$, the persistence of a partially suppressed N\'eel order up to a higher field $\mu_\mathrm{0}H_\mathrm{p}$ does not agree with the expectation for a continuous QPT in a single XXZ spin-$\frac{1}{2}$ chain \cite{Yang_1966,Haldane_1980,Bogoliubov_1986}.

\begin{figure}[t]
\centering
	\includegraphics[width=0.49\textwidth]{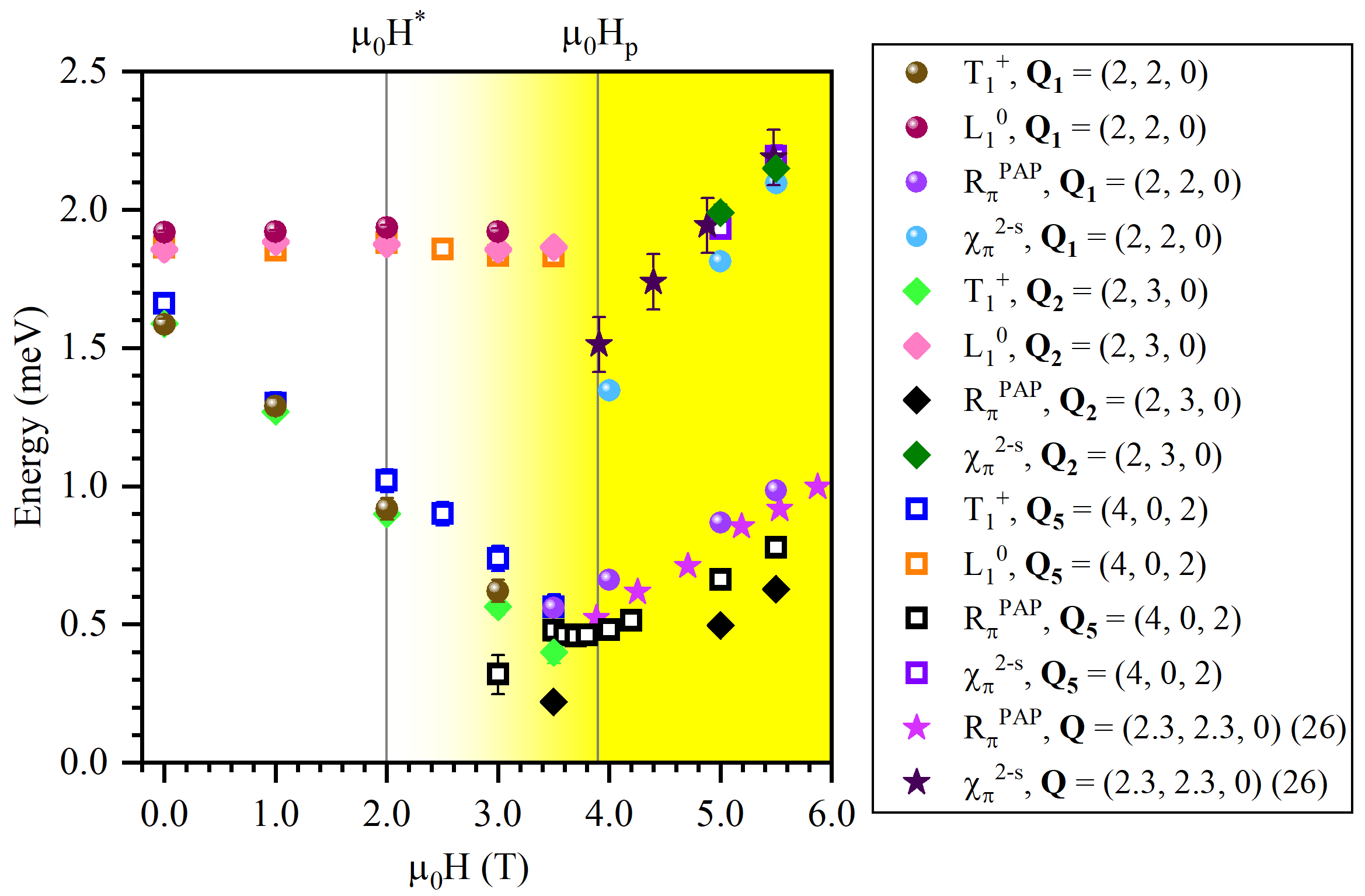}
    \caption{Evolution of the quasiparticle excitation energies in a longitudinal magnetic field. $\mathbf{Q_{1}}$ and $\mathbf{Q_{5}}$ are nuclear zone centers; $\mathbf{Q_{2}}$ is an antiferromagnetic zone center. For comparison, the data reported in Ref.~\cite{Bera_2020} are also included.}
\label{fig:5}
\end{figure}

To confirm the nature of the magnetic state between $\mu_\mathrm{0}H^{\star}$ and $\mu_\mathrm{0}H_\mathrm{p}$, we now consider the changes seen in the INS spectra as a function of magnetic field at 50~mK. A full description of the TLL spin dynamics is now available in both theory and experiment: they consist of multiparticle excitations ranging from psinon-antipsinon and psinon-psinon pairs at low energies, to Bethe strings at medium and high energies \cite{Wang_2018,Faure_2019,Bera_2020,Yang_2019}, as illustrated in Fig.~\ref{fig:3}(a-d). While these modes can still be effectively described by spin flips, i.e.~having a quantum spin number +1, -1 or 0 [see Fig.~\ref{fig:3}(b-d)], they differ from the spinons in the N\'eel state in that they correspond to the excitations in a disordered spin-$\frac{1}{2}$ chain with $M_{c}$~$\neq$~0. Figure \ref{fig:4}(a) shows the INS spectrum in the (2, K, 0) direction at 5.5 T, i.e. above $\mu_\mathrm{0}H_\mathrm{p}$.  All INS spectra in this region can be modelled by combining two DHOs and a product of a Lorentzian and Heaviside step function [Fig.~\ref{fig:4}(b)].  By comparing our data with the canonical observations of the TLL spin dynamics, the dominant features in the spectrum can be identified. The 4-fold screw chain structure in SrCo$_{2}$V$_{2}$O$_{8}$ folds the Brillouin zone by a factor of four, meaning that the excitations at L~=~0, $\frac{1}{4}$, $\frac{1}{2}$ and $\frac{3}{4}$ of a single chain can be observed simultaneously at L~=~0 \cite{Bera_2020}. Most of the magnetic excitations in a TLL contribute to a continuum, accounting for the broad Lorentzian-Heaviside profile \cite{Gannon_2019}. There are only two relatively well defined modes that are accessible at L~=~0 and below 2.5~meV: the psinon-antipsinon pairs at L~=~$\frac{1}{2}$ ($\mathrm{R_{\pi}^{PAP}}$) with a quantum number $S_\mathrm{PAP}$~=~0 and Bethe 2-strings at L~=~$\frac{1}{2}$ ($\chi_{\pi}^{2-s}$) with a quantum number $S_\mathrm{2-s}$~=~-1. By comparing the field dependences of the two DHO centers with those reported in Ref.~\cite{Bera_2020}, we confirm that the profile below 1.5~meV is the $\mathrm{R_{\pi}^{PAP}}$ mode, while the one around 2.1~meV is the $\chi_{\pi}^{2-s}$ mode (Fig.~\ref{fig:5}). Our in-plane geometry is optimal for detecting modes with a quantum number of 0, which is consistent with the observed intense spectral weight for the $S_\mathrm{PAP}$~=~0 $\mathrm{R_{\pi}^{PAP}}$ mode. A remarkable feature revealed in Fig.~\ref{fig:4}(a) is the dispersion of this mode; it is strongly softened at the AFM zone centers with a bandwidth of 0.36(1)~meV, contrasting sharply with the dispersionless $\chi_{\pi}^{2-s}$ mode at higher energies. This strongly suggests that the interchain couplings must be taken into account to understand the low-energy TLL spin dynamics.

\begin{figure*}
\centering
	\includegraphics[width=0.9\textwidth]{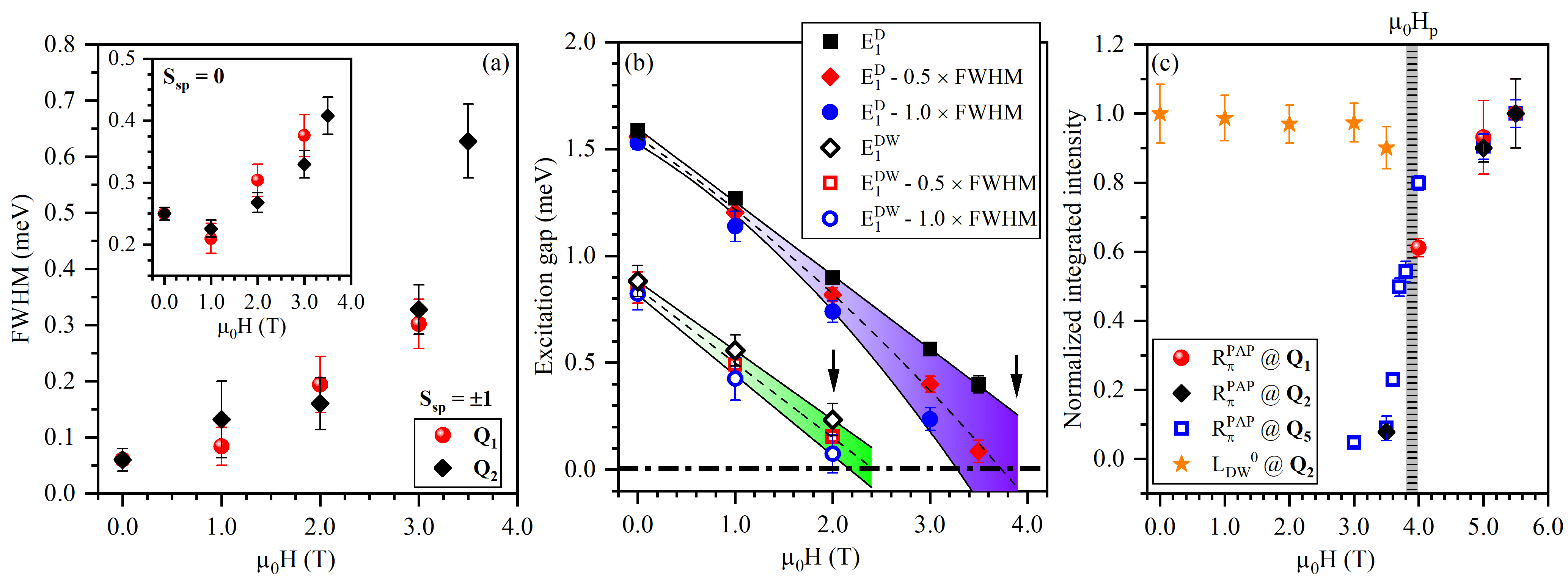}
    \caption{Spinon damping and spatially inhomogeneous gap closure. (a) Intrinsic full width at half maximums (FWHMs) of the $S_\mathrm{sp}$~=~$\pm$1 (main panel) and $S_\mathrm{sp}$~=~0 (inset) modes as a function of magnetic field. The instrumental resolution contribution has been deconvolved. $\mathbf{Q_{1}}$~=~(2,~2,~0) and $\mathbf{Q_{5}}$~=~(4,~0,~2) are nuclear zone centers; $\mathbf{Q_{2}}$~=~(2,~3,~0) is an antiferromagnetic zone center. (b) Estimated spinon excitation gaps in the domains (filled points) and domain walls (open points) at $\mathbf{Q_{2}}$ as a function of magnetic field. $\mu_\mathrm{0}H^{\star}$ and $\mu_\mathrm{0}H_{p}$ are marked by black arrows. The shaded areas are guides to the eye. (c) Magnetic field dependences of the normalized integrated intensity of the $\mathrm{R_{\pi}^{PAP}}$ and $\mathrm{L_\mathrm{DW}^{0}}$ modes. The hatched column marks the position of $\mu_\mathrm{0}H_{p}$. All measurements were performed at 50~mK.}
\label{fig:6}
\end{figure*}

With the features of the relevant TLL modes established, we now discuss the region between $\mu_\mathrm{0}H^{\star}$ and $\mu_\mathrm{0}H_\mathrm{p}$. In Figure~\ref{fig:4}(c), we present the magnetic excitations and their dispersion at 3.5~T, where the elastic intensity at $\mathbf{Q_{2}}$ that characterizes the N\'eel order is at about 83 $\%$ of its zero-field value [Fig.~\ref{fig:3}(f)]. A visual inspection already reveals two strong modes: a weakly dispersive one around 1.85~meV and a strongly dispersive one with a bandwidth of 0.31(3)~meV below 0.6~meV. The former is identical with the $\mathrm{L_{1}^{0}}$ spinon bound state within the errors [Fig.~\ref{fig:2}(c)], revealing the persistence of the spinon confinement in the bulk N\'eel domains. The latter, however, cannot be explained by any $S_\mathrm{sp}$~=~+1 mode in the system, which is dispersionless [Fig.~\ref{fig:2}(c)] and approaches the elastic line as the magnetic field increases by gaining Zeeman energy. In fact, these spectra can be numerically reproduced by a minimal model with four modes: $\mathrm{L_{1}^{0}}$, $\mathrm{L_\mathrm{DW}^{0}}$, $\mathrm{T_{1}^{+}}$, and, critically, a dispersive $\mathrm{R_{\pi}^{PAP}}$. In these analyses, the contributions from the elastic line, which are not negligible at low energies, have been subtracted by fitting the zero-field spectrum. A typical example is demonstrated in Fig.~\ref{fig:4}(d) for the spectrum at $\mathbf{Q_{2}}$, while we have also validated this model at $\mathbf{Q_{1}}$ and another nuclear zone center $\mathbf{Q_{5}}$~=~(4,~0,~2) (Fig.~\ref{fig:5}). The existence of a strongly dispersive mode can be resolved down to 3.0~T in our measurements. By evaluating its evolution as a function of magnetic field, the mode energy seamlessly connects to the S$_\mathrm{PAP}$~=~0 $\mathrm{R_{\pi}^{PAP}}$ one above $\mu_\mathrm{0}H_{p}$ (Fig.~\ref{fig:5}). The field dependence and strong dispersion of the new mode below $\mu_\mathrm{0}H_{p}$ do not fit the characteristics of spinons, but are in strong agreement with the features of a $\mathrm{R_{\pi}^{PAP}}$. Consequently, the most likely explanation for our observations at these intermediate magnetic fields is N\'eel\,/\,TLL coexistence, with the TLL component accounting for the partial suppression of the bulk N\'eel order revealed by neutron diffraction.

\subsection{Inhomogeneous N\'eel to Tomonaga-Luttinger liquid quantum phase transition}

In Section \ref{sec:3}, we have argued that the chains in and close to the domain walls may have different excitation gaps from those in the domains based on a minimal model. This possibility is supported by the zero-field INS data presented in Section \ref{sec:4}~A, which show the presence of a second set of spinon excitations with a smaller gap. We have interpreted these results as the fingerprints of excitations localized inside the domain walls. Now, we use this interpretation to address the N\'eel\,/\,TLL coexistence phenomenon shown in Section \ref{sec:4}~B.

As noted above, the transition to the TLL state can be related to the condensation of the relevant $\mathrm{T_{1}^{+}}$ mode, brought about by Zeeman splitting.  Figure \ref{fig:5} illustrates that we do not see a clear closure of the gap associated with the $\mathrm{T_{1}^{+}}$ mode in the bulk of the domains. As noted earlier, the spinon confinement is not perfect in this material, and the intrinsic damping of the spinon bound states can be extracted from the FWHMs after deconvolving the instrumental resolution.  The resulting spinon damping is shown in Fig.~\ref{fig:6}(a), and is enhanced in both transverse and longitudinal modes by the applied magnetic field.  This finite bandwidth will lead to a smaller gap for at least part of the excitations. We have calculated the excitation gap in the domains using three different criteria: $E\mathrm{_{1}^{D}}$, $E\mathrm{_{1}^{D}}$~-~(0.5~$\times$~FWHM) and $E\mathrm{_{1}^{D}}$~-~(1.0~$\times$~FWHM), where $E\mathrm{_{1}^{D}}$ is the central energy of the $\mathrm{T_{1}^{+}}$ mode at a given magnetic field. Their magnetic field dependences are plotted in Fig.~\ref{fig:6}(b). As in the case of the charge density wave in in NbSe$_3$ \cite{NbSe3} we speculate that once sufficient spectral weight accumulates at zero energy transfer, condensation occurs.  It is only by including this magnetic-field-induced spinon damping that this can be achieved at $\mu_\mathrm{0}H_\mathrm{p}$ for the $\mathrm{T_{1}^{+}}$ mode originating from the domains.

The same methodology can be applied to the transverse spinon bound state in the domain walls, the central energy of which is denoted as $\mathrm{E_{1}^{DW}}$ in Fig.~\ref{fig:6}(b). While $\mathrm{E_{1}^{DW}}$ can be unambiguously determined at zero-field (Fig.~\ref{fig:2}), we were not able to directly measure its value once the magnetic field is switched on due to its marginal spectral weight. But we can infer its behaviour by invoking the general Zeeman splitting principle: $\mathrm{E_{1}^{DW}}$~($\mu_\mathrm{0}H$)~=~$\mathrm{E_{1}^{DW}}$~(0~T)~-~$g_{c}\mu_\mathrm{B}\mu_\mathrm{0}H$. As shown in Fig.~\ref{fig:5}, this principle is well obeyed by the other measurable spinon bound states at higher energies. A linear fit to the $\mathrm{T_{1}^{+}}$ mode energies at different $\mathbf{Q}$s gives $g_{c}$~=~5.6(2). By further including the spinon damping effect, we find that the excitation gap of the $\mathrm{T_{DW}^{+}}$ band at $\mathbf{Q_{2}}$ closes near $\mu_\mathrm{0}H^{\star}$ [Figs.~\ref{fig:6}(b)], above which $M_{c}$ becomes non-zero [Fig.~\ref{fig:3}(e)]. Combining the elastic and inelastic neutron scattering results, we believe that the $\mathrm{R_{\pi}^{PAP}}$ mode, which can be resolved down to 3.0~T within our detection limit, originates from the local gap closure in the domain walls near $\mu_\mathrm{0}H^{\star}$.

The spatially inhomogeneous QPT scenario is further backed up by the behavior of the spectral weight of the $\mathrm{R_{\pi}^{PAP}}$ mode [Fig.~\ref{fig:6}(c)]. At and below 3.5~T, it evolves smoothly and remains small. This is because the TLL at these intermediate fields is only located inside and close to the domain walls. When the excitation gap in the N\'eel domain begins to close, or the magnetic field approaches $\mu_\mathrm{0}H_\mathrm{p}$ [Fig.~\ref{fig:5}(b)], the spectral weight shows a step-like increase, suggesting the emergence of a N\'eel to TLL QPT in the domains. The N\'eel order is completely suppressed at high fields, where the spectral weight is flattened again.

\section{\label{sec:5}Discussion and Conclusions}

The results above support an exotic region between $\mu_\mathrm{0}H^{\star}$ and $\mu_\mathrm{0}H_\mathrm{p}$ where a partially suppressed N\'eel state coexists with the TLL. One possible mechanism for this is masked quantum criticality. This means that the N\'eel to TLL QPT in SrCo$_{2}$V$_{2}$O$_{8}$ is no longer well defined. In correlated electron systems, masked quantum criticality can occur due to the development of a competing state near the QCP \cite{Mathur_1998,Yuan_2003}. However, after an extensive search in the reciprocal space, no emergent static spin modulation can be resolved in this region in addition to the N\'eel order (see Fig.~\ref{fig:S3} in the Supplemental Material). As shown in Fig.~\ref{fig:4}(c)~$\&$~(d), the dynamical spin properties in the $\mathbf{Q}$-E space can also be well described by the N\'eel / TLL composite model. Our data therefore do not support the competing state scenario.

The persistence of the $\mathrm{L_\mathrm{DW}^{0}}$ mode up to $\mu_\mathrm{0}H_\mathrm{p}$ [Figs.~\ref{fig:1}(g)~$\&$~\ref{fig:6}(c)] would seem to disagree with the domain wall QPT scenario at $\mu_\mathrm{0}H^{\star}$.  This is due to the lack of significant interactions between the different spinon bound states in the low field region, where these states are relatively stable. At high fields, spinon damping becomes profound [Fig.~\ref{fig:6}(a)], promoting these interactions, such as the decay of a $S_\mathrm{sp}$~=~0 mode into a pair of $S_\mathrm{sp}$~=~$\pm$1 modes \cite{Lake_2000}. In fact, this echoes our argument in Section~\ref{sec:4} C that sufficient spinon damping is necessary for the N\'eel to TLL QPT to percolate.

In summary, we have presented neutron scattering evidence for inhomogeneous spin excitations in the the quasi 1D quantum magnet SrCo$_{2}$V$_{2}$O$_{8}$, both in the zero-field N\'eel ground state and during the field-induced N\'eel to TLL QPT process. What is particularly interesting is the coexistence of partially suppressed N\'eel antiferromagnetism and a TLL between $\mu_\mathrm{0}H^{\star}$~=~2.0~T and $\mu_\mathrm{0}H_\mathrm{p}$~=~3.9~T, which supports the hypothesis of masked quantum criticality in this system. Masked quantum criticality is an important concept in correlated electron systems, contributing to some of the most intricate states of matter observed therein \cite{Mathur_1998,Yuan_2003,Pfleiderer_2004,Uemura_2007,Friedemann_2018,Broun_2008,Shibauchi_2014}. Spatial confinement, i.e. dimensionality control, of particles is a well-established tool for tuning the quantum phases; paradigmatic examples can be found in systems like ultracold atoms \cite{Mun_2007,Chomaz_2016} and superfluid $^{4}$He in porous media \cite{Reppy_1992}. Unlike these studies where the confining potential is extrinsic, our work highlights an intrinsic mechanism for confinement, i.e. dimensional crossover, that can be used to explore the quantum magnetism in relevant materials.

\vspace*{0.3cm}

\begin{acknowledgments}
LS, EC and EB acknowledge financial support from the Swedish Research Council under project no. 2018-04704. D.~P.~and A.~T.~B.~acknowledge the UK EPSRC for funding under grant number EP/N034872/1. Z.~H.~thanks the National Natural Science Foundation of China (NSFC) (No.~U1632159 and No.~21875249) for financial support. The authors gratefully acknowledge the ILL for the allocated beamtime. The data collected at the ILL are available \cite{ILLDATA1,ILLDATA2,ILLDATA3}. Part of this work is based on experiments performed at the Swiss spallation neutron source SINQ, Paul Scherrer Institute, Villigen, Switzerland. 
\end{acknowledgments}

\section*{Supplemental Material}

In the Supplemental Material, the SrCo2V2O8 samples grown by floating zone \cite{Lejay_2011} and spontaneous nucleation \cite{He_2006} are labelled as SCVO-FZ and SCVO-SN, respectively.

\subsection*{Magnetic field versus temperature phase diagram}

\begin{figure}[h]
	\centering
	\includegraphics[width=0.49\textwidth]{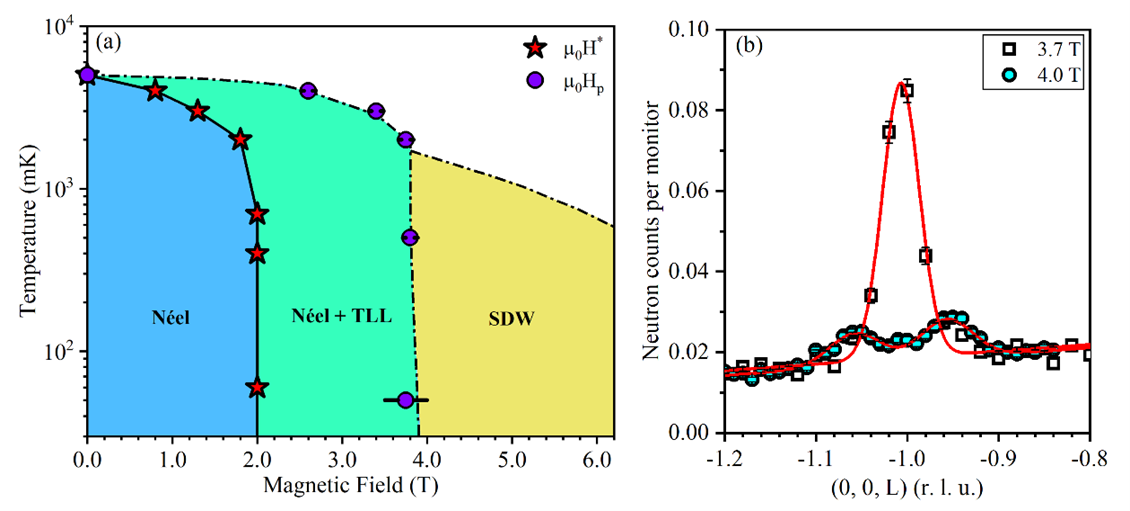}
	\caption{(a) Magnetic field ($H // c$ axis) versus temperature phase diagram. The dash-dotted lines are from measurements on SCVO-SN and are reported in Ref.~\cite{Shen_2019}. The solid circles and stars are measurements on SCVO-FZ. $\mu_0 H^*$ and $\mu_0 H_p$ represent the magnetic fields where the gap closes in the domain walls and domains, respectively. Below about 1.5 K, a spin density wave (SDW) appears above $\mu_0 H_p$ , splitting the commensurate peak belonging to the Néel order into a pair of incommensurate peaks along the reciprocal $\bm{c^*}$ axis, as shown in (b) for SCVO-FZ.}
	\label{fig:S1}
\end{figure}

\subsection*{Magnetic structure refinement between $\mu_0 H^*$ and $\mu_0 H_p$}

\begin{figure}[h]
	\centering
	\includegraphics[width=0.49\textwidth]{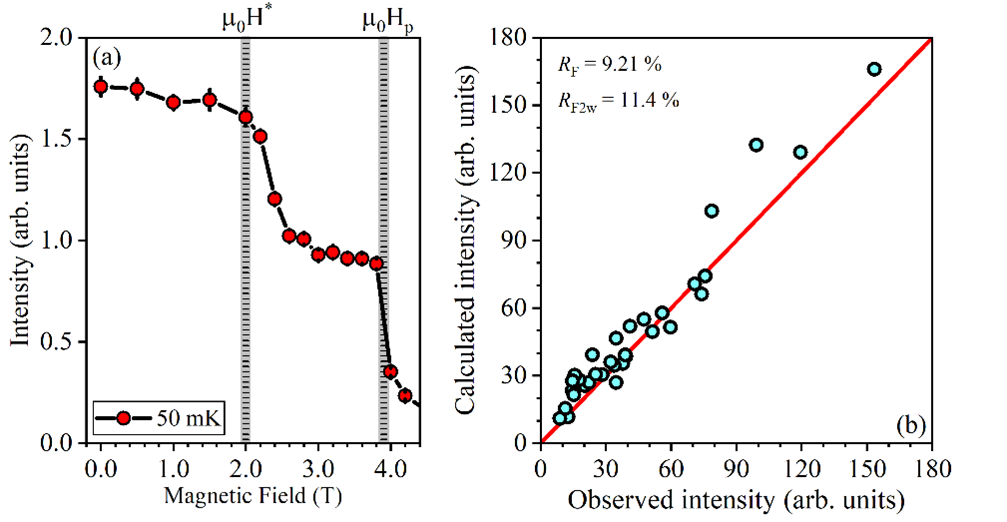}
	\caption{(a) Magnetic field dependence of the intensity at $\bm{Q}_2$ = (2, 3, 0) at 50 mK. The gray hatched columns mark the positions of $\mu_0 H^*$ and $\mu_0 H_p$. (b) Magnetic structure refinement on the N\'eel phase at $T$ = 50 mK and $\mu_0 H$ = 3.0 T. The R-factors ($R_F$, $R_{F2w}$) which characterize the quality of our refinement are also given. The data were taken on SCVO-SN at ZEBRA (PSI).}
	\label{fig:S2}
\end{figure}

Due to the small crystal size (see Methods section), only neutron diffraction measurements have been performed on SCVO-SN. The magnetic field dependence of the diffraction intensity at $\bm{Q}_2$ = (2, 3, 0) and at 50 mK is shown in Fig.~\ref{fig:S2}(a). In this sample, the Bragg reflection is also partially suppressed above 2.0 T ($\mu_0 H^*$), but the suppression appears to be much sharper. Between 2.5 T and $\mu_0 H_p$= 3.9 T, a plateau is seen. The distinct field responses near $\mu_0 H^*$ between the two samples could be caused by differences in sample quality. We note that no impurity phase can be detected in either of these two samples.

To check the magnetic structure here, we collected 13 nuclear reflections and 41 magnetic reflections belonging to the N\'eel phase at 3.0 T and 50 mK. These measurements were made in the same experiment as the magnetic structure refinements published in Ref.~\cite{Shen_2019}, measured at 0.0 T and 50 mK, using the same instrumental configuration. As a result, the refined magnetic structure at 3.0 T can be directly compared to that at 0.0 T. The refinement was carried out with the FULLPROF package, using the procedure described in detail in Ref.~\cite{Shen_2019}. In Figure \ref{fig:S2}(b), we have plotted the correlation curve between the experimental and calculated intensities of these magnetic reflections. Compared to the structure in zero field, the only field-induced change revealed in our refinement is that the ordered magnetic moment per Co drops from 1.81(4)47 $\mathrm{\mu_B}$ to 1.32(4) $\mathrm{\mu_B}$, while the volume fractions of the two magnetic domains in this compound remain unchanged relative to those measured in zero-field.

\subsection*{Search for emergent static spin modulation between $\mu_0 H^*$ and $\mu_0 H_p$}

\begin{figure}[h]
	\centering
	\includegraphics[width=0.49\textwidth]{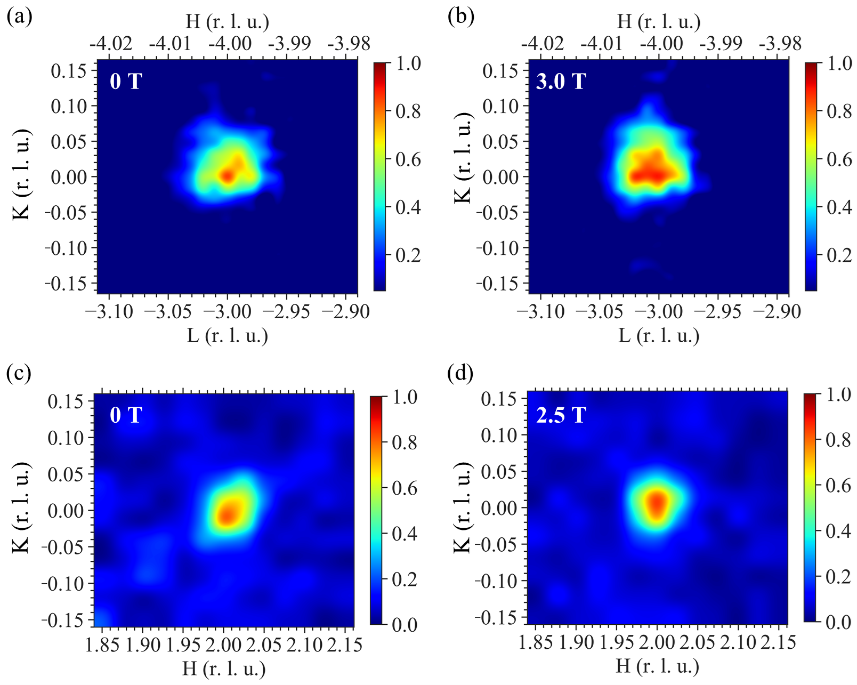}
	\caption{Reciprocal space search using neutron diffraction. (a) \& (b) Selected neutron diffraction patterns measured around (-4, 0, -3) at 0 T and 3.0 T. (c) \& (d) Selected neutron diffraction patterns measured around (2, 0, 0) at 0 T and 2.5 T. All measurements were performed at 30 mK. The data were taken on SCVO-SN at D10 (ILL).}
	\label{fig:S3}
\end{figure}

One possible scenario for explaining the masked quantum criticality in SCVO is the appearance of competing states \cite{Mathur_1998, Yuan_2003}. The area detector on the D10 diffractometer at the ILL allowed us to do an extensive search for an emergent static spin modulation over broader regions of reciprocal space in SCVO-SN. Around the N\'eel antiferromagnetic zone center (-4, 0, -3), we have searched in the region with $\Delta H = \pm 0.05$, $\Delta K = \pm 0.44$ and $\Delta L = \pm 0.35$ r.l.u.; no emergent spin modulation can be detected at $T$ = 30 mK and 2.0 T $\le \mu_0 H \le$  3.9 T. Around the nuclear zone center (2, 0, 0), we have searched in the H-K plane with $\Delta H =  \pm 0.9$ and $\Delta K =  \pm 0.9$ r.l.u.; again no emergent incommensurate modulation can be found within our resolution. Figure \ref{fig:S3} shows some of the data resulting from this search, concentrating on the regions close to the zone centers. These results are consistent with our study on SCVO-FZ [Fig.~\ref{fig:S1}b].

\end{document}